\documentclass[journal]{IEEEtran}

\usepackage[OT1]{fontenc} 
\usepackage[numbers,sort&compress]{natbib}
\usepackage[cmex10]{amsmath}
\usepackage{amssymb}
\usepackage{bm}
\usepackage{braket}
\usepackage{graphicx}
\usepackage{color}
\usepackage{subfigure}
\usepackage{tabularx}
\usepackage{arydshln}
\usepackage{mathtools}
\usepackage{multirow}
\usepackage{algorithm,algorithmic}
\usepackage{url}
\usepackage{listings}
\usepackage{comment}
\usepackage{CJKutf8}
\usepackage{here}

\graphicspath{{./figure/}}

\begin{document}
\title{\LARGE Quantum Speedup for Polar Maximum Likelihood Decoding}

\author{Shintaro~Fujiwara,~\IEEEmembership{Graduate Student Member,~IEEE} and Naoki~Ishikawa,~\IEEEmembership{Senior~Member,~IEEE}.\thanks{S.~Fujiwara and N.~Ishikawa are with the Faculty of Engineering, Yokohama National University, 240-8501 Kanagawa, Japan (e-mail: fujiwara-shintaro-by@ynu.jp). This research was partially supported by the Japan Society
for the Promotion of Science (JSPS) KAKENHI (Grant Numbers 23K22755).}}

\markboth{\today}
{Shell \MakeLowercase{\textit{et al.}}: Bare Demo of IEEEtran.cls for Journals}
\maketitle

\begin{abstract}
Conventional decoding algorithms for polar codes strive to balance achievable performance and computational complexity in classical computing.
While maximum likelihood (ML) decoding guarantees optimal performance, its NP-hard nature makes it impractical for real-world systems.
In this letter, we propose a novel ML decoding architecture for polar codes based on the Grover adaptive search, a quantum exhaustive search algorithm. 
Unlike conventional studies, our approach, enabled by a newly formulated objective function, uniquely supports Gray-coded multi-level modulation without expanding the search space size compared to the classical ML decoding.
Simulation results demonstrate that our proposed quantum decoding achieves ML performance while providing a pure quadratic speedup in query complexity.
\end{abstract}


\IEEEpeerreviewmaketitle

\section{Introduction}
\IEEEPARstart{P}{olar} codes, introduced by Arikan \cite{arikan2009channel}, are the first class of error-correcting codes capable of achieving Shannon capacity over binary input discrete memoryless symmetric channels with low encoding complexity, which is adopted in the 5G standard for uplink control channels. The original decoding algorithm for polar codes, the successive cancellation (SC) algorithm, operates with $O(N\log N)$ time and space complexity, where $N$ is the code length. 
However, the SC decoding suffers from suboptimal error-correction performances, especially for short to moderate code lengths, and cannot compete with turbo and low-density parity check (LDPC) codes.

To address this limitation, the successive cancellation list (SCL) decoding algorithm was developed~\cite{tal2015list}, retaining the $L$ most likely codewords at each step to enhance error-correction capability. Integrating cyclic redundancy check codes with SCL decoding further improves the performance~\cite{tal2015list}, often surpassing turbo and LDPC codes for short to moderate code lengths. However, this approach increases complexity to $O(LN \log N)$. When $L = 2^K$, covering all possible polar codewords, it achieves optimal maximum likelihood (ML) decoding, but with exponential complexity.

In quantum computing, Grover's algorithm~\cite{grover1996fast} can locate an exact solution in an unsorted database of size $2^N$ with $O(\sqrt{2^N})$ query complexity, offering a quadratic speedup over the $O(2^N)$ complexity of classical exhaustive search. This framework has been extended to binary optimization problems while retaining the quadratic speedup. Recent studies have explored Grover adaptive search (GAS) \cite{gilliam2021grover}, a variant of Grover's algorithm, to reduce complexity in solving NP-hard problems like codebook design~\cite{yukiyoshi2024quantum} and ML decoding for multi-input multi-output systems~\cite{norimoto2023quantum}.\footnote{Note that such speedup can only be realized through future fault-tolerant quantum computing.}

Several studies have explored decoding algorithms for classical polar codes relying on quantum computing. A pioneering approach of \cite{kasi2022decoding} uses a quantum gate model with amplitude amplification, a generalization of Grover's algorithm, to achieve ML decoding under ideal conditions with binary phase shift keying (BPSK) modulation. However, this approach does not consider practical multi-level modulation schemes and evaluates all possible $2^N$ binary vectors, including non-polar codewords, whereas the classical ML decoding only searches $2^K$ polar codewords. Another approach of \cite{kasi2024quantum} combines quantum annealing with classical computation, where the classical part computes log-likelihood ratios at the upper-level of the SC decoder, and the quantum part decodes lower-level sub-blocks. The study in \cite{seah2023xSA} shows this annealing approach without upper-level classical computation can achieve near-ML performance for short polar codes. However, it requires redundant ancillae, expanding the search space to $2^{N(\log N + 1)}$.
Additionally, quantum annealing often produces sub-optimal solutions, and achieving quantum speedup over classical computing is generally considered unlikely \cite{Rønnow2014Quantum}.

Against this background, we consider achieving a pure quantum speedup for polar ML decoding. This letter has two major contributions.
1) \textbf{We propose an ML decoding architecture for classical polar codes using GAS, achieving a pure quadratic speedup in computational complexity over a classical ML counterpart.} Specifically, the classical ML decoding requires $O(2^K)$ complexity, while our approach requires only $O(\sqrt{2^K})$.
This speedup relies on an initial state preparation exploiting the unique structure of polar codes, which prepares a superposition of quantum states corresponding to only valid polar codewords.
2) Additionally, \textbf{we introduce the first approach for ML decoding of classical polar codes with a multi-level Gray-coded $2^M$-pulse amplitude modulation~(PAM).} This approach features a tailored objective function for ML decoding of multi-level Gray-coded symbols, which is also applicable to general ML detection schemes, while 
the conventional studies \cite{kasi2024quantum,kasi2022decoding, seah2023xSA} only support BPSK.

\section{Background}
\label{sec:background}

\subsection{Polar Codes}
We consider a symmetric discrete memoryless channel with input alphabet $\mathcal{X}=\{0,1\}$ and output alphabet $\mathcal{Y}$. With $N$ representing the codeword length, polar encoding is expressed as $\mathbf{x}_{0}^{N-1} = \mathbf{u}_{0}^{N-1}\mathbf{G}_{N}$, where $\mathbf{x}_{0}^{N-1} = [x_0,\cdots,x_{N-1}]\in \mathbb{F}_{2}^{N}$ denotes a codeword vector, $\mathbf{u}_{0}^{N-1} = [u_0,\cdots,u_{N-1}]\in \mathbb{F}_{2}^{N}$ denotes an information vector, and $\mathbf{G}_N \in \mathbb{F}_{2}^{N\times N}$ denotes a generator matrix.
The generator matrix is formulated as 
\begin{equation}
\mathbf{G}_N = \mathbf{G}_2^{\otimes n} \text{ with } \ \mathbf{G}_2 = 
\begin{bmatrix}
1 & 0 \\
1 & 1 \\
\end{bmatrix},
\end{equation}
where $n = \log_2 N$ and $\mathbf{G}_2^{\otimes n}$ is the $n$-th Kronecker power of $\mathbf{G}_2$. 
An input sequence of polar codes is consist of $K$ information bits and $N-K$ frozen bits, where the position of these bits are determined relying on the reliabilities of bit channels.
We denote $\mathcal{A}$ and $\mathcal{F}$ as the sets of information bit indices and frozen bit indices, respectively.
In addition, when the multi-level Gray-coded $2^M$-PAM symbols are adopted, the $s$-th $(0\leq s\leq M-1)$ sequence of information bits and codeword bits are expressed as $\mathbf{u}_{s} = [u_{s,0},u_{s,1},\cdots, u_{s,{N-1}}]$ and  $\mathbf{x}_{s} = [x_{s,0},x_{s,1},\cdots, x_{s,{N-1}}]$, respectively.

\subsection{GAS Algorithm}
GAS \cite{gilliam2021grover} is a quantum algorithm designed to solve combinatorial optimization problems involving $N$ binary variables, extending Grover's algorithm.  
It offers a quadratic speedup in query complexity over classical exhaustive search: while the classical approach requires $O(2^N)$ complexity to find a solution among $2^N$ candidates, GAS reduces this to $O(\sqrt{2^N})$.

Let $E(\mathbf{x})$ be an objective function to minimize, and assume its minimum and maximum values satisfy the inequalities concerning a natural number $m$ as $-2^{m - 1} \leq \min_{\mathbf{x}} E(\mathbf{x})$ and $\max_{\mathbf{x}} E(\mathbf{x}) < 2^{m - 1}$.
Suppose such an objective function is given, the quantum circuit for GAS requires $N$ qubits to represent a binary variable vector $\mathbf{x}$, and $m$ qubits to represent the objective function values. The former qubits are referred to as {\em key part} and the latter are referred to as {\em value part}.

In the GAS algorithm, firstly, an initial quantum state is prepared as a superposition of all possible quantum states by applying Hadamard gates $\mathbf{H}^{\otimes N}$ to $\ket{0}^{\otimes N}$. 
Secondly, a quantum dictionary subroutine \cite{gilliam2021foundational} encodes an objective function value $E(\mathbf{x}) - c_i$, where $c_i$ is a tentative threshold value for the $i$-th iteration of the GAS algorithm. 
The objective function values are represented in two's complement form, 
allowing the most significant qubit in the value part to express the sign of the values.
The above operations are expressed as an operator $\mathbf{A}_{c_i}$. 
Thirdly, an oracle $\mathbf{O}$ consisting of a single Z gate flips the phase of the state if that qubit is $\ket{1}$.
Finally, the Grover diffusion operator $\mathbf{D}$ \cite{grover1996fast} amplifies the amplitude of the states with objective function values less than the threshold value $c_i$.
A Grover operator is expressed as
$\mathbf{G} = \mathbf{A}_{c_{i}} \mathbf{D} \mathbf{A}_{c_{i}}^{\mathrm{H}} \mathbf{O}$,
and the final state of the GAS algorithm is given by
\begin{equation}
    \ket{\psi} = \mathbf{G}^{L_i} \mathbf{A}_{c_i}\ket{0}_{N+m},
\end{equation}
where $L_i$ is the number of iterations of the GAS algorithm and the optimal iteration number can be approximated as
\begin{equation}
    L_{\mathrm{opt}} = \left\lfloor \frac{\pi}{4} \sqrt{2^N} \right\rfloor,
\end{equation}
when there is only one solution.
The entire GAS algorithm is summarized in Algorithm~\ref{alg:gas-algo}. 

To evaluate the complexity of the GAS algorithm, two metrics defined in \cite{botsinis2014fixedcomplexity} have been commonly used: the total numbers of objective function evaluations in classical domain~(CD), i.e., $i$ at the end of the algorithm, and Grover operators $\mathbf{G}$ applied in quantum domain~(QD), i.e., $L_0 + L_1 + \cdots + L_i$, which is termed query complexity.

\begin{algorithm}[t]
    \caption{GAS designed for real-valued coefficients \cite{gilliam2021grover,norimoto2023quantum}.}\label{alg:gas-algo}
    \begin{algorithmic}[1]
        \renewcommand{\algorithmicrequire}{\textbf{Input:}}
        \renewcommand{\algorithmicensure}{\textbf{Output:}}
        \REQUIRE Objective function $E:\mathbb{B}^N\rightarrow\mathbb{R}$ and $\lambda=8/7$
        \ENSURE A solution $\mathbf{x} \in \mathbb{B}^{N}$
        \STATE {Uniformly sample $\mathbf{x}_0 \in \mathbb{B}^N $ and set $c_0=E(\mathbf{x}_0)$}. 
        \STATE {Set $k = 1$ and $i = 0$}.
        \REPEAT
        \STATE\hspace{\algorithmicindent}{Sample a rotation count $L_i$ from $\{0, 1, \cdots, \lceil k-1 \rceil$\}}.
        \STATE\hspace{\algorithmicindent}{Evaluate $\mathbf{G}^{L_i} \mathbf{A}_{y_i} \Ket{0}_{N+m}$, and obtain $\mathbf{x}$}. 
        \STATE\hspace{\algorithmicindent}{Evaluate $c=E(\mathbf{x})$ on a classical computer}. 
        \hspace{\algorithmicindent}\IF{$c<c_i$}
        \STATE\hspace{\algorithmicindent}{$\mathbf{x}_{i+1}=\mathbf{x}, c_{i+1}=c,$ and $k=1$}.
        \hspace{\algorithmicindent}\ELSE{\STATE\hspace{\algorithmicindent}{$\mathbf{x}_{i+1}=\mathbf{x}_i, c_{i+1}=c_i,$ and $k=\min{\{\lambda k,\sqrt{2^N}}\}$}}.
        \ENDIF
        \STATE{$i=i+1$}.
        \UNTIL{a termination condition is met}.
    \end{algorithmic} 
\end{algorithm}

\begin{figure*}[t]
\includegraphics[trim={0.8cm 0.5cm 1cm 0.5cm},scale=0.8]{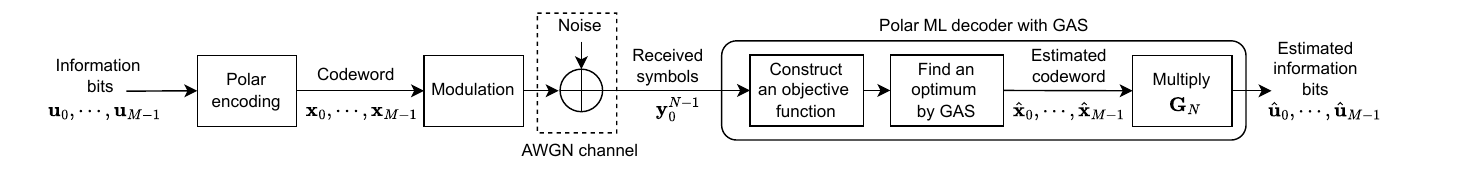}
\centering
\caption{System model of our proposed ML decoding architecture with GAS for polar codes.}
\label{fig:system_model}
\end{figure*}

\subsection{State-of-the-Art Approach}
We introduce the approach of \cite{kasi2024quantum}, which decodes polar codes using a hybrid method combining quantum annealing and classical computing.
The quantum annealing decodes lower-level sub-blocks of an SC decoder, formulated as a quadratic unconstrained binary optimization (QUBO) problem. The decoded information vector is represented as a binary vector $\hat{\mathbf{b}} \in \mathbb{B}^N$ which minimizes a QUBO objective function.

The objective function for the formulation is defined as a sum of cost functions for an encoding constraint $C_\mathrm{E}$, a frozen constraint $C_\mathrm{F}$, and a receiver constraint $C_\mathrm{R}$ in \cite{kasi2024quantum, seah2023xSA}, i.e.,
\begin{equation}
\begin{aligned}\label{eq:kasi_qubo}
    \hat{\mathbf{b}} = \underset{\mathbf{b}}{\arg\min}\left\{ W_\mathrm{E} \sum_{\forall (i,j) \in \mathcal{T}} C_\mathrm{E}(b_i,b_j) + W_\mathrm{F} \sum_{\forall i \in \mathcal{F}} C_\mathrm{F}(b_i) \right. \\
    \left. + W_\mathrm{R} \sum_{\forall i \in \mathrm{EEC}} C_\mathrm{R}(b_i) \right\}.
\end{aligned}
\end{equation}
Here, $W_\mathrm{E}$, $W_\mathrm{F}$, and $W_\mathrm{R}$ are penalty coefficients for each constraint. In \cite{kasi2024quantum}, the suggested optimal parameters are $W_\mathrm{E} = 1$, $W_\mathrm{F} = 4$, and $W_\mathrm{R} = 2-R_{\mathrm{sub}}$, where $R_{\mathrm{sub}}$ is the code rate of the sub-block being decoded.

In \eqref{eq:kasi_qubo}, the encoding constraint ensures the correct operation of polar encoding and is defined as
\begin{equation}
    C_\mathrm{E}(b_i,b_j) = (b_i + b_j - a_k - 2a_{k+1})^2
\end{equation}
for $\forall(i,j)\in \mathcal{T}$, where $\mathcal{T}$ represents all possible pairs of input qubit indices for the XOR operations in the $\mathbf{G}_2$ operations. 
Throughout the polar encoding, XOR operations are performed $(N/2) \log_2 N$ times in total, and two auxiliary binary variables $a_k, a_{k+1}$ are required per XOR operation. As a result, $N \log_2 N$ ancilla qubits are necessary for this constraint.

Meanwhile, the frozen constraint ensures that the frozen bits are fixed to 0 and is defined as
\begin{equation}
    C_\mathrm{F}(b_i) = b_i.
\end{equation}

Finally, the receiver constraint $C_\mathrm{R}$ ensures that the equivalent encoded codeword (EEC), which corresponds to the information bits at the highest level of the sub-block, matches the information of received codewords.

The formulation \eqref{eq:kasi_qubo} requires $N (\log_2 N + 1)$ binary variables, and the search space size increases to $2^{N (\log_2 N + 1)}$, while that of the classical ML decoding is $2^{K}$. Therefore, the ML decoding with this approach is practically infeasible.  
Additionally, in \cite{kasi2024quantum}, the decoding performance of this approach was shown only with BPSK modulation, and the extension for multi-level modulation has not been discussed. 

\section{Proposed Approach}
\label{sec:prop}

\subsection{Quantum Polar ML Decoding Architecture}
The system model of our proposed polar ML decoding architecture using GAS is shown in Fig.~\ref{fig:system_model}.

At the transmitter, the process begins by encoding the information bit vectors $\mathbf{u}_0, \mathbf{u}_1, \cdots, \mathbf{u}_{M-1}$, producing codeword vectors $\mathbf{x}_s = \mathbf{u}_s\mathbf{G}_N$ for $s=0, 1, \cdots, M-1$. These codewords are then modulated into symbols $\mathcal{M}(\mathbf{z}_i)$ for $i = 0,1, \cdots, N-1$, where $\mathbf{z}_{i} = [x_{0,i}, x_{1,i}, \cdots, x_{M-1,i}]$. Here, BPSK or Gray-coded $2^M$-PAM is applied, after which the symbols are transmitted over an additive white Gaussian noise (AWGN) channel.

At the receiver, the received symbol vector $\mathbf{y}_{0}^{N-1} = [y_{0}, y_{1}, \cdots, y_{N-1}]$ is input into the ML decoder with GAS, which minimizes an objective function based on the received vector to estimate codewords $\hat{\mathbf{x}}_{s}$ with query complexity $O(\sqrt{2^{MK}})$, especially $O(\sqrt{2^{K}})$ when the BPSK is applied. Finally, the estimated information bits $\hat{\mathbf{u}}_{s}$ are obtained by multiplying $\hat{\mathbf{x}}_{s}$ by $\mathbf{G}_N$, since $\mathbf{G}_N = \mathbf{G}_N^{-1}$ in $\mathbb{F}_2$.

In this letter, BPSK and $2^M$-PAM schemes are assumed, which only consider symbol distribution on the real axis. Note that, however, our approach can be easily extended to modulation schemes such as quadrature amplitude modulation~(QAM), which includes distribution on the imaginary axis.

\subsection{Proposed Initial State Preparation and Objective Functions}
In the proposed ML decoding method, the initial state is prepared as a superposition of all possible polar codewords. 
In polar coding, frozen bits are fixed to 0, while information bits take values of 0 or 1. Thus, qubits corresponding to information bits are expressed as an equal superposition of $\ket{0}$ and $\ket{1}$, while qubits for frozen bits are fixed to $\ket{0}$. The superposition of $\ket{0}$ and $\ket{1}$ is created by applying a Hadamard gate to $\ket{0}$, i.e., $\mathbf{H}\ket{0} = (1/\sqrt{2})(\ket{0} + \ket{1})$.

Subsequently, by applying appropriate circuit operations for polar encoding to the superposition, the initial state can be prepared as a superposition of
all possible polar codewords.
As a result, the search space size is reduced, and the encoding and frozen bit constraints in \eqref{eq:kasi_qubo} can be omitted, simplifying the objective function to focus solely on the receiver constraint.

We define the objective function as the distance between the estimated and received symbol vectors:
\begin{equation}
    E(\mathbf{x}) = \sum_{i=0}^{N-1} |y_{i} -\mathcal{M}(\mathbf{z}_i)|^2.
\end{equation}

The following describes methods for preparing the initial state and deriving objective functions tailored for BPSK and the general Gray-coded $2^M$-PAM.

\subsubsection{BPSK Modulation}
In classical polar encoding, $\mathbf{G}_2$ operation acts on two input bits, leaving one unchanged and setting the other as the result of their XOR. A general operation $\mathbf{G}_N$ is constructed by recursively applying the $\mathbf{G}_2$ operations.

In a quantum circuit, the $\mathbf{G}_2$ operation can be implemented with a CNOT gate \cite{kasi2022decoding}, where the control qubit remains unchanged, and the target qubit is set as the XOR result. The $\mathbf{G}_N$ operation can similarly be implemented using CNOT gates, following the classical polar encoder structure, allowing the preparation of a superposition of valid polar codewords. By restricting the search space to valid codewords, the space size reduces to $2^K$, matching that of the ML decoding with the classical exhaustive search. This encoding circuit requires $(N/2)\log_2 N$ CNOT gates and a circuit depth of $O(\log N)$.

\begin{figure}[!t]
 \centering
 \includegraphics[trim={2cm 0.5cm 2cm 0.5cm},scale=0.65]{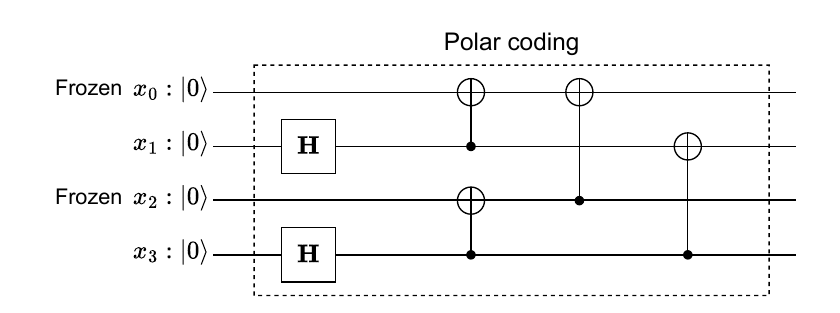}
 \caption{Quantum circuit for preparing a initial state for BPSK modulation.}
 \label{fig:initial_state}
\vspace{-.4cm}
\end{figure}

Fig.~\ref{fig:initial_state} examplifies the quantum circuit for preparing the initial state for $(N,K)=(4,2)$ and $\mathcal{F}=\{0,2\}$. In this case, Hadamard gates are applied only to the first and third qubits, and the circuit requires 4 CNOT gates with a depth of 2.

The objective function for BPSK symbols is defined as the distance between received and estimated symbol vectors:
\begin{equation}
    E(\mathbf{x}) =  \sum_{i=0}^{N-1} C_\mathrm{R}(x_i)  =  \sum_{i=0}^{N-1}\left|y_i-(1-2x_i) \right|^2,
\end{equation}
which is simplified using $x_i^2 = x_i$ and removing a constant:
\begin{equation}\label{eq:simplifed}
    E'(\mathbf{x}) = \sum_{i=0}^{N-1} y_ix_i.
\end{equation}
All terms in \eqref{eq:simplifed} are first-order, providing an implementation advantage for the quantum circuit for GAS.

\subsubsection{Gray-Coded $2^M$-PAM}
Next, we consider the scenario where the codeword bits are modulated as the Gray-coded $2^M$-PAM $(M \geq 2)$ symbols, where adjacent symbols differ by one bit, and all bit levels have the same reliability. In this setup, we assume that the information bits are encoded using a bit-interleaved coded modulation (BICM) approach~\cite{seidl2013multilevel}. Specifically, an information bit sequence of length $MN$ is divided into $M$ sequences of $N$ bits and each of sequence is encoded by $\mathbf{G}_N$ operation. An interleaver optionally shuffles the $MN$ codeword bits. In quantum computing, this interleaver can be implemented using SWAP gates that swap the quantum states of two qubits.

In the classical decoding, each polar codeword modulated by the BICM method is decoded in parallel by its corresponding decoders, which leads to suboptimal decoding performance. However, in this letter, we consider an ML decoder that jointly estimates all codeword vectors based on all received symbols.

The objective function for $2^M$-PAM is expressed as the distance between the received and estimated symbol vectors. The coordinate on the real axis of the $i$-th estimated symbol corresponding to $\mathbf{z}_{i} = [x_{0,i}, x_{1,i}, \cdots, x_{M-1,i}]$ is given by
\begin{equation}
\begin{split}
    S_{\mathrm{G}}(\mathbf{z}_{i}) 
    = \frac{1}{\sqrt{A}} \sum_{j=0}^{M-1} 2^{M-j-1} (-1)^{j} \left( \prod_{k=0}^{j} (1 - 2x_{k,i}) \right),
\end{split}
\end{equation}
where $A$ is a scaling factor. Therefore, the objective function for Gray-coded $2^M$-PAM symbols is written as
\begin{align}\label{eq:objfun_wo_encoding}
    E(\mathbf{x}) = \sum_{i=0}^{N-1} \left| y_i - S_{\mathrm{G}}(\mathbf{z}_{i}) \right|^2,
\end{align}
which results in a higher-order unconstrained binary optimization (HUBO) formulation of $M$-th order. The number of terms in this function grows exponentially with respect to $M$, potentially leading to inefficiencies when implementing a quantum circuit for GAS.

To reduce the order and number of terms in \eqref{eq:objfun_wo_encoding}, we use the classical differential encoding and decoding techniques dating back to Gray's patent in 1947. 
Differential encoding translates natural labeling symbols into Gray-coded symbols as
\begin{equation}
    x'_{s,i} = 
    \begin{cases}
        x_{s,i} & \text{if } s = 0, \\
        x_{s-1,i} \oplus x_{s,i} & \text{if } s = 1,2,\cdots,M-1,
    \end{cases}
    \label{eq:differential_bit_encoding}
\end{equation}
where $x'_{s,i}$ is the encoded bit after differential encoding. The inverse operation, differential decoding, translates Gray-coded symbols back into natural labeling symbols as
\begin{equation}
    x_{s,i} = \bigoplus_{j=0}^{s} x'_{j,i}.
    \label{eq:differential_bit_encoding_reverse}
\end{equation}
After applying differential decoding, the coordinate of the estimated symbol corresponding to the codeword bits $\mathbf{z}'_i = (x'_{0,i}, \cdots, x'_{M-1,i})$ is given by
\begin{equation}
\begin{split}
    S_{\mathrm{N}}(\mathbf{z}'_i) 
    = \frac{1}{\sqrt{A}} \sum_{j=0}^{M-1} 2^{M-j-1} (-1)^{j} (1 - 2x'_{j,i}).
\end{split}
\end{equation}
Now, a new objective function in a QUBO form is written as
\begin{align}\label{eq:objfun_w_encoding}
    E(\mathbf{x}) = \sum_{i=0}^{N-1} \left| y_i - S_{\mathrm{N}}(\mathbf{z}'_i) \right|^2.
\end{align}

\begin{figure}[!t]
 \centering
 \includegraphics[trim={2cm 0.5cm 2cm 0.5cm},scale=0.65]{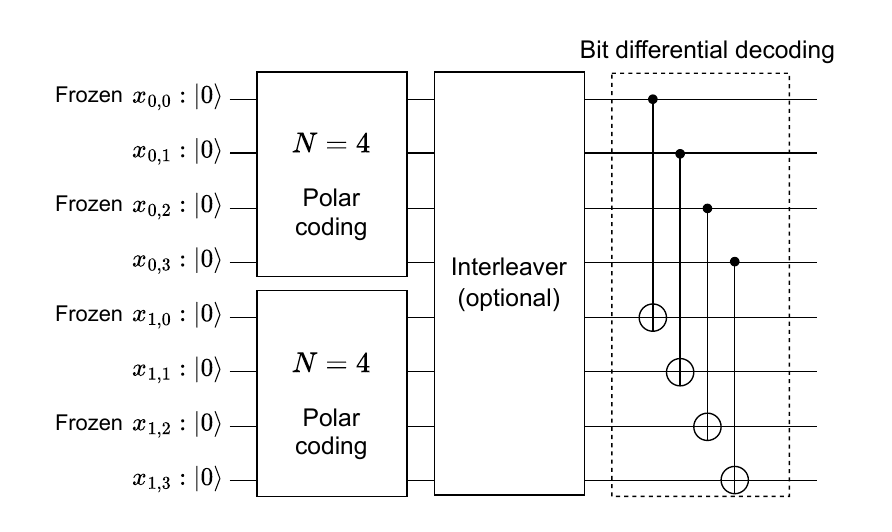}
 \caption{Quantum circuit for preparing the initial state for $4$-PAM.}
 \label{fig:BICM_initial_state}
\vspace{-.4cm}
\end{figure}

Fig.~\ref{fig:BICM_initial_state} exemplifies the quantum circuit to prepare the initial state for polar code of $(N,K)=(4,2)$ with $\mathcal{F}=\{0,2\}$, where 4-PAM and differential encoding is adopted. Note that the quantum circuit for differential encoding can be omitted when using the objective function \eqref{eq:objfun_wo_encoding} instead of \eqref{eq:objfun_w_encoding}.

Note that the objective functions \eqref{eq:objfun_wo_encoding} and \eqref{eq:objfun_w_encoding} can be applied not only for the polar ML decoding but the general ML detection problem of multi-level modulation symbols.

\subsection{Modified Uniform Sampling}
In the original GAS algorithm, one of the $2^{MN}$ possible quantum states is randomly selected to calculate the initial threshold $c_0$ in the classical domain. In contrast, our method prepares the initial state as a superposition of only $2^{MK}$ states corresponding to valid polar codewords. Selecting a state outside this superposition may result in a smaller objective function value, leading to decoding failure.

To address this issue, we modify the algorithm to randomly sample only from the $2^{MK}$ valid states that satisfies the frozen bit constraint. The sampled sequence is then encoded to generate the corresponding codewords. This process requires $O(MN\log N)$ complexity, with additional complexity for interleaving and bit differential encoding in the classical domain. However, these operations are performed only once at the beginning of the algorithm, which does not affect the overall complexity.

\section{Simulation Results}

\begin{figure}[t]
 \centering
 \includegraphics[width=1.0\hsize]{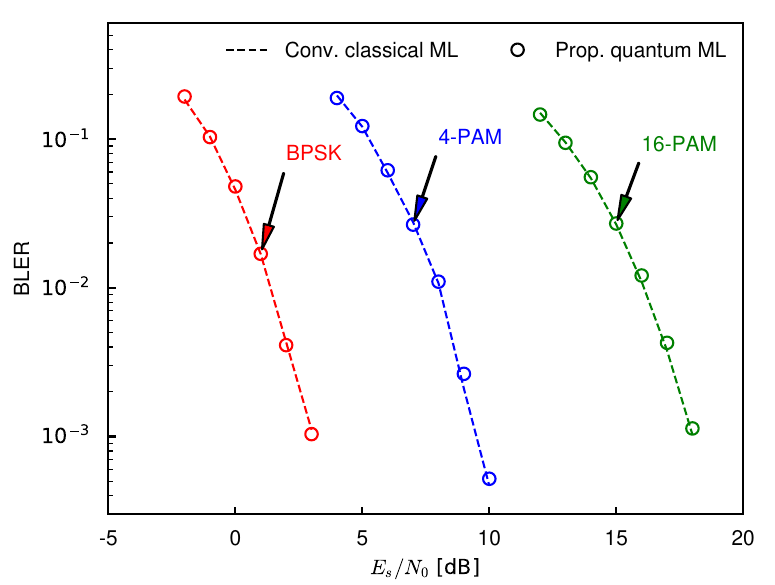}
 \caption{BLER performance comparisons of the conventional exhaustive search and the proposed quantum ML decoding.}
 \label{fig:BLER_comparison}
\vspace{-.4cm}
\end{figure}

\label{sec:sim}

In this section, the performance of the proposed polar ML decoder with GAS is evaluated in terms of block error rate (BLER) and query complexity.

Firstly, Fig.~\ref{fig:BLER_comparison} presents the achievable BLER performance of the ML decoding using both classical exhaustive search and GAS, which is equivalent to the quantum exhaustive search. 
Simulations were performed under three scenarios: $(N,K)=(16,8)$ with $\mathcal{F} = \{0, 1, 2, 3, 4, 5, 6, 8\}$ and BPSK; $(N,K)=(8,4)$ with $ \mathcal{F} = \{0, 1, 2, 4\}$ and $4$-PAM; and $(N,K)=(4,2)$ with  $\mathcal{F} = \{0, 2\}$ and $16$-PAM. 
As Fig.~\ref{fig:BLER_comparison} indicates, both classical and quantum ML decoding achieved equivalent error performance, validating the reliability of the proposed quantum approach.

\begin{figure}[ht]
 \centering
 \includegraphics[width=1.0\hsize]{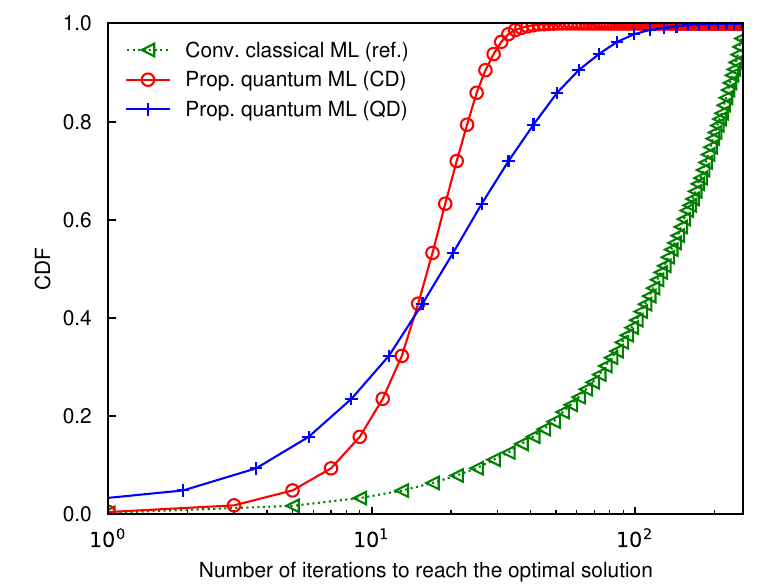}
 \caption{CDF of the number of iterations required to reach an optimal solution with ML decoding when $(N,K)=(16,8)$ and BPSK modulation is applied.}
 \label{fig:meanobj_comparison_BPSK}
\vspace{-.4cm}
\end{figure}

\begin{figure}[ht]
 \centering
 \includegraphics[width=1.0\hsize]{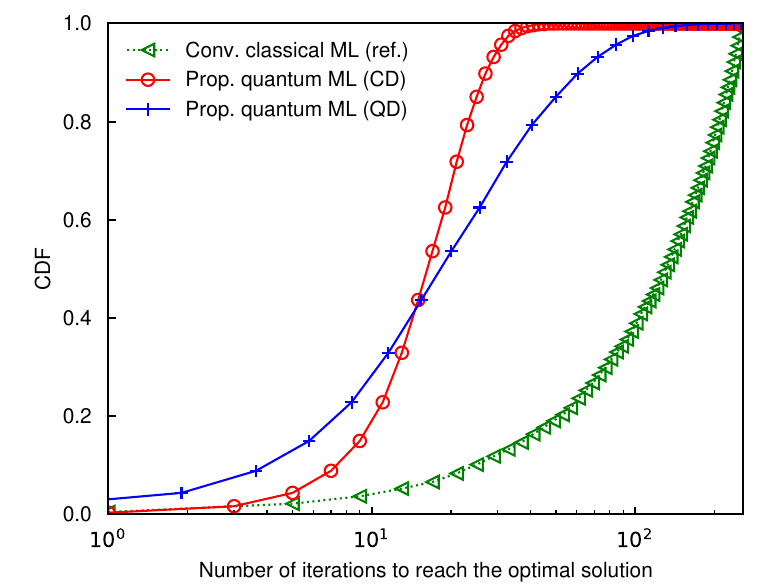}
 \caption{CDF of the number of iterations required to reach an optimal solution with ML decoding when $(N,K)=(4,2)$ and $16$-PAM is applied.}
 \label{fig:meanobj_comparison_16PAM}
\vspace{-.4cm}
\end{figure}

Secondly, Figs.~\ref{fig:meanobj_comparison_BPSK} and~\ref{fig:meanobj_comparison_16PAM} show the cumulative distribution function (CDF) of the number of iterations required to reach the optimal solution for two cases: $N=16$ with BPSK and $N=4$ with $16$-PAM. 
As shown, the proposed quantum ML decoding exhibited significant speedups in query complexity over the classical ML decoding.
The size of search space with our approach was just $2^8$ for both cases, which was equivalent to that of the classical ML. In contrast, if the conventional formulation \cite{kasi2024quantum, seah2023xSA} in \eqref{eq:kasi_qubo} were used, the search space size would increase to $2^{90}$ and $2^{48}$ for the first and second cases, respectively.
Therefore, it is clear that the conventional formulation of \eqref{eq:kasi_qubo} is not suitable for the ML decoding, even with GAS.

\section{Conclusion}
In this letter, we proposed a polar ML decoding architecture using GAS, supporting multi-level Gray-coded $2^M$-PAM. This is the only approach that does not expand the search space size in the context of quantum-assisted polar ML decoding.
Simulations showed that our approach achieved a pure quadratic speedup in decoding complexity over the classical counterpart, while maintaining the optimal performance. 
Future work includes extending the approach to other error-correcting codes.

\footnotesize{
	\bibliographystyle{IEEEtranURLandMonthDiactivated}
	\bibliography{main}

\begin{thebibliography}{10}
\providecommand{\url}[1]{#1}
\csname url@samestyle\endcsname
\providecommand{\newblock}{\relax}
\providecommand{\bibinfo}[2]{#2}
\providecommand{\BIBentrySTDinterwordspacing}{\spaceskip=0pt\relax}
\providecommand{\BIBentryALTinterwordstretchfactor}{4}
\providecommand{\BIBentryALTinterwordspacing}{\spaceskip=\fontdimen2\font plus
\BIBentryALTinterwordstretchfactor\fontdimen3\font minus \fontdimen4\font\relax}
\providecommand{\BIBforeignlanguage}[2]{{%
\expandafter\ifx\csname l@#1\endcsname\relax
\typeout{** WARNING: IEEEtran.bst: No hyphenation pattern has been}%
\typeout{** loaded for the language `#1'. Using the pattern for}%
\typeout{** the default language instead.}%
\else
\language=\csname l@#1\endcsname
\fi
#2}}
\providecommand{\BIBdecl}{\relax}
\BIBdecl

\bibitem{arikan2009channel}
E.~Ar{\i}kan, ``Channel polarization: A method for constructing capacity-achieving codes for symmetric binary-input memoryless channels,'' \emph{IEEE Trans. Inform. Theory}, vol.~55, no.~7, pp. 3051--3073, 2009.

\bibitem{tal2015list}
I.~Tal and A.~Vardy, ``List decoding of polar codes,'' \emph{IEEE Trans. Inform. Theory}, vol.~61, no.~5, pp. 2213--2226, 2015.

\bibitem{grover1996fast}
L.~K. Grover, ``A fast quantum mechanical algorithm for database search,'' in \emph{{{ACM}} Symp. on {{Theory}} of Comput.}, Philadelphia, PA, USA, 1996, pp. 212--219.

\bibitem{gilliam2021grover}
A.~Gilliam, S.~Woerner, and C.~Gonciulea, ``Grover adaptive search for constrained polynomial binary optimization,'' \emph{Quantum}, vol.~5, p. 428, 2021.

\bibitem{yukiyoshi2024quantum}
K.~Yukiyoshi, T.~Mikuriya, H.~Rou, G.~de~Abreu, and N.~Ishikawa, ``Quantum speedup of the dispersion and codebook design problems,'' \emph{IEEE Trans. Quantum Eng.}, vol.~5, no.~01, pp. 1--16, 2024.

\bibitem{norimoto2023quantum}
M.~Norimoto, R.~Mori, and N.~Ishikawa, ``Quantum algorithm for higher-order unconstrained binary optimization and {{MIMO}} maximum likelihood detection,'' \emph{IEEE Trans. Commun.}, vol.~71, no.~4, pp. 1926--1939, 2023.

\bibitem{kasi2022decoding}
S.~Kasi, J.~Kaewell, S.~Hamidi-Rad, and K.~Jamieson, ``Decoding polar codes via noisy quantum gates: Quantum circuits and insights,'' \emph{arXiv:2210.10854}, 2022.

\bibitem{kasi2024quantum}
S.~Kasi, J.~Kaewell, and K.~Jamieson, ``A quantum annealer-enabled decoder and hardware topology for next{G} wireless polar codes,'' \emph{IEEE Trans. Wirel. Commun.}, vol.~23, no.~4, pp. 3780--3794, 2024.

\bibitem{seah2023xSA}
R.~Seah, H.~Zhou, M.~Jalaleddine, and W.~J. Gross, ``x{SA}: A binary cross-entropy simulated annealing polar decoder,'' in \emph{International Symposium on Topics in Coding}, 2023, pp. 1--5.

\bibitem{Rønnow2014Quantum}
T.~F. Rønnow, Z.~Wang, J.~Job, S.~Boixo, S.~V. Isakov, D.~Wecker, J.~M. Martinis, D.~A. Lidar, and M.~Troyer, ``Defining and detecting quantum speedup,'' \emph{Science}, vol. 345, no. 6195, pp. 420--424, 2014.

\bibitem{gilliam2021foundational}
A.~Gilliam, C.~Venci, S.~Muralidharan, V.~Dorum, E.~May, R.~Narasimhan, and C.~Gonciulea, ``Foundational patterns for efficient quantum computing,'' \emph{arXiv:1907.11513}, 2021.

\bibitem{botsinis2014fixedcomplexity}
P.~Botsinis, S.~X. Ng, and L.~Hanzo, ``Fixed-complexity quantum-assisted multi-user detection for {{CDMA}} and {{SDMA}},'' \emph{IEEE Trans. Commun.}, vol.~62, no.~3, pp. 990--1000, 2014.

\bibitem{seidl2013multilevel}
M.~Seidl, A.~Schenk, C.~Stierstorfer, and J.~B. Huber, ``Multilevel polar-coded modulation,'' in \emph{IEEE Int. Symp. Inf. Theory}, 2013, pp. 1302--1306.

\end{thebibliography}
}

\end{document}